\documentclass[aps,prl,twocolumn,10pt,showpacs,superscriptaddress,amsmath,amssymb,nobibnotes]{revtex4-1}

\usepackage{graphicx}
\usepackage{amsfonts}    
\usepackage{graphicx}   
\usepackage{verbatim}   
\usepackage{color}      
\usepackage{subfigure}  
\usepackage{hyperref}   
\usepackage{natbib}
\usepackage{booktabs}
\usepackage{threeparttable}
\usepackage{dcolumn}
\usepackage{multirow}
\usepackage{float}
\usepackage{ulem}
\usepackage{bm}
\usepackage{algorithm}
\usepackage{algorithmicx}
\usepackage[noend]{algpseudocode}

\makeatletter
\def\BState{\State\hskip-\ALG@thistlm}
\makeatother
\begin{document}


\title{Efficient Classical Computation of Quantum Mean Values for Shallow QAOA Circuits }

\author{Wei-Feng Zhuang}
\thanks{These authors contributed equally to this work.}
\affiliation{Beijing Academy of Quantum Information Sciences, Beijing 100193, China}

\author{Ya-Nan Pu}
\thanks{These authors contributed equally to this work.}
\affiliation{Beijing Academy of Quantum Information Sciences, Beijing 100193, China}

\author{Hong-Ze Xu}
\affiliation{Beijing Academy of Quantum Information Sciences, Beijing 100193, China}

\author{Xudan Chai}
\affiliation{Beijing Academy of Quantum Information Sciences, Beijing 100193, China}

\author{Yanwu Gu}
\affiliation{Beijing Academy of Quantum Information Sciences, Beijing 100193, China}

\author{Yunheng Ma}
\affiliation{Beijing Academy of Quantum Information Sciences, Beijing 100193, China}

\author{Shahid Qamar}
\affiliation{Beijing Academy of Quantum Information Sciences, Beijing 100193, China}

\author{Chen Qian}
\affiliation{Beijing Academy of Quantum Information Sciences, Beijing 100193, China}

\author{Peng Qian}
\affiliation{Beijing Academy of Quantum Information Sciences, Beijing 100193, China}

\author{Xiao Xiao}
\affiliation{Beijing Academy of Quantum Information Sciences, Beijing 100193, China}

\author{Meng-Jun Hu}
\email{humj@baqis.ac.cn}
\affiliation{Beijing Academy of Quantum Information Sciences, Beijing 100193, China}

\author{Dong E. Liu}
\email{dongeliu@mail.tsinghua.edu.cn}
\affiliation{State Key Laboratory of Low Dimensional Quantum Physics, Department of Physics, Tsinghua University, Beijing, 100084, China}
\affiliation{Beijing Academy of Quantum Information Sciences, Beijing 100193, China}
\affiliation{Frontier Science Center for Quantum Information, Beijing 100184, China}

\date{\today}

\begin{abstract}

The Quantum Approximate Optimization Algorithm (QAOA), which is a variational quantum algorithm, aims to give sub-optimal solutions of combinatorial optimization problems. It is widely believed that QAOA has the potential to demonstrate application-level quantum advantages in the noisy intermediate-scale quantum(NISQ) processors with shallow circuit depth. Since the core of QAOA is the computation of expectation values of the problem Hamiltonian, an important 
practical question is whether we can find an efficient classical algorithm to solve quantum mean value in the case of general shallow quantum circuits. Here, we present a novel graph decomposition based classical algorithm that scales linearly with the number of qubits for the shallow QAOA circuits in most optimization problems except for complete graph case. Numerical tests in Max-cut, graph coloring and Sherrington-Kirkpatrick model problems, compared to the state-of-the-art method, shows orders of magnitude performance improvement. Our results are not only important for the exploration of quantum advantages with QAOA, but also useful for the benchmarking of NISQ processors.       
\end{abstract}

\maketitle



\section{Introduction}
The rapid development of quantum computing technologies in past decades has attracted a lot of interests from both academia and industrial community. In 2019, Google demonstrated the so-called ``quantum supremacy" by using the 53- qubit superconducting quantum processor {\it Sycamore} \cite{sycamore}. Although the Google's claim seems controversial now \cite{claim1, claim2, claim3}, stronger quantum advantage on the same problem has been verified by the USTC team with a higher quality 60-qubit superconducting quantum processor {\it Zuchongzhi}~\cite{zu1, zu2}. These works imply that we have arrived the so-called noisy intermediate-scale quantum (NISQ) era \cite{NISQ}, where the quantum processors contain about fifty to a few hundreds noisy qubits. Quantum error correction and fault tolerance could in principle support the quantum computation with arbitrary accuracy, but on the other hand, they require a much larger scale processor with an error rate in each physical element mitigated to an extremely low level. Unfortunately, due to high error rate of qubit operations, those NISQ processors are limited to shallow depth circuit without error corrections. In that case, a naturally important question arises: whether or not application-level quantum advantage can be demonstrated in NISQ processors.

Variational quantum algorithms (VQA), which includes variational quantum eigensolver (VQE), quantum machine learning (QML) and quantum approximate optimization algorithm (QAOA), has been widely believed as a promising approach for demonstrating NISQ applications, such as quantum chemistry \cite{chem1, chem2, chem3}, machine learning \cite{ml1, ml2, ml3} and combinatorial optimization \cite{op1, op2, op3}. The basic idea of VQA is to estimate the cost function by sampling from parameterized quantum circuit and calls for classical optimizer to find iterative parameters until convergence condition is reached \cite{VQA}. The bitstring as solution is obtained in the last stage by sampling from quantum circuit with the optimalized parameters. The cost function is in general expectation value of a Hamiltonian that can be written as a linear combination of poly$(n)$ Pauli operators, e.g., Ising Hamiltonian in the QAOA. Since the quantum circuit sampling has been shown classically intractable as the number of qubits increase \cite{sampling1, sampling2, sampling3}, it is natural to ask whether or not quantum mean value of shallow depth circuit can be calculated efficiently by a classical computer. If the answer is positive, then NISQ processor is only necessary in the quantum sampling part of QAOA.

\begin{figure*}[tbp]
\centering
\includegraphics[scale=0.5]{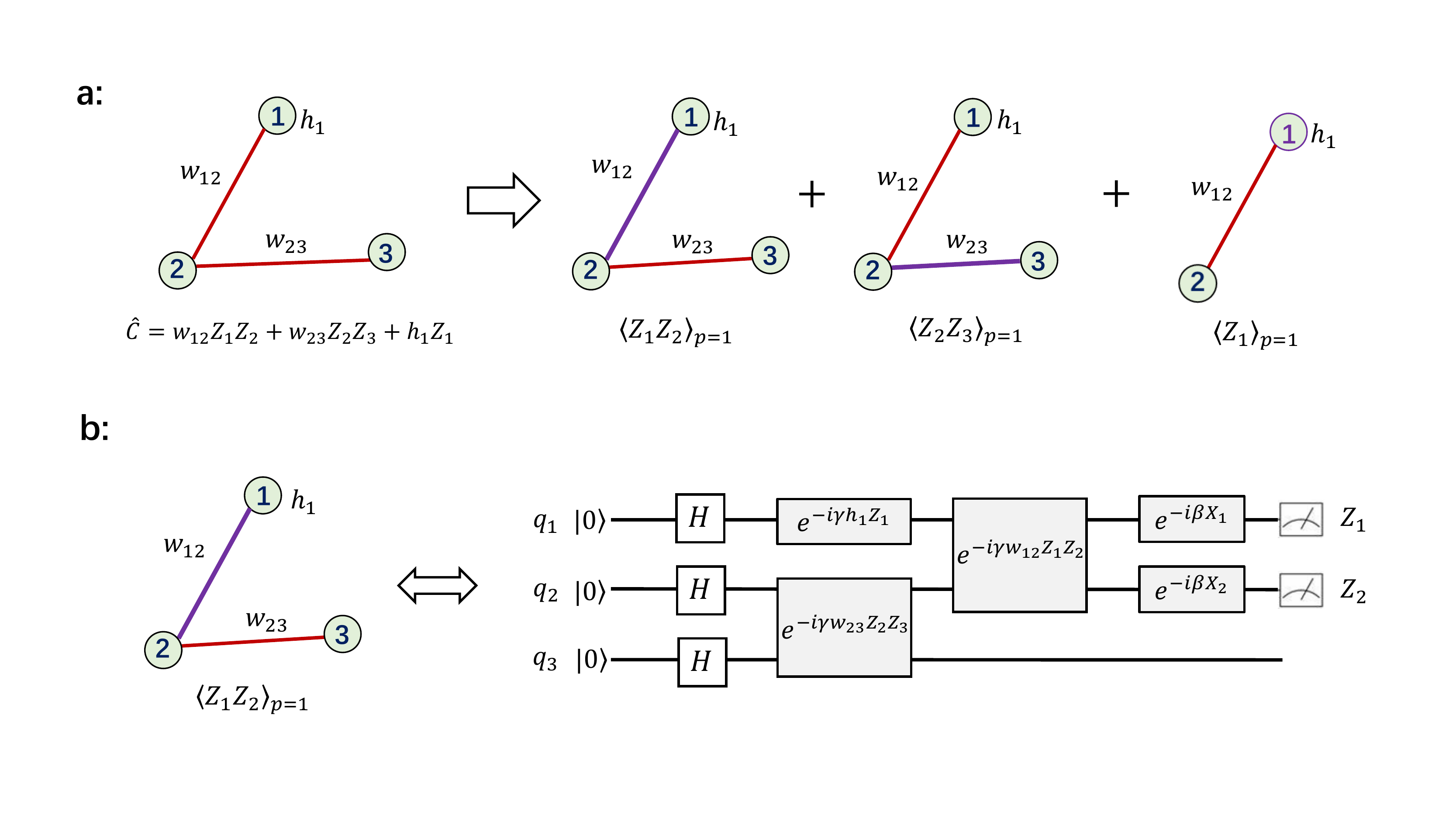}
\caption{Example illustration of basic procedure of graph decomposition algorithm. \textbf{a:} The cost operator $\hat{C}$ is mapped to a weighed graph $G$, in which $w_{ij}$ and $h_{k}$ are the weight of edge $(ij)$ and bias of node $k$ respectively. According to the value of $p$, we only show $p=1$ case here, the G can be decomposed into independent subgraphs that correspond to terms $\langle\hat{Z}_{i}\hat{Z}_{j}\rangle_{p}$ and $\langle\hat{Z}_{k}\rangle_{p}$. The number of subgraphs equals to the number of individual terms in $\hat{C}$. \textbf{b:} There is an one-to-one correspondence between the subgraphs and the quantum circuits for computation of $\langle\hat{Z}_{i}\hat{Z}_{j}\rangle_{p}$ and $\langle\hat{Z}_{k}\rangle_{p}$. Detail realization is given in {\bf Algorithm 1}.}
\label{f1}
\end{figure*} 

In a recent paper, Bravyi, Gosset and Movassagh show that there exist the classical algorithms scales linearly with the number of qubits for the special case of geometrically local two-dimensional quantum circuits \cite{mean}. Whether the quantum mean value problems can be solved efficiently on a classical computer in the case of general shallow circuits, however, remains a central open question. Here, we show that there does exist a classical algorithm capable of efficiently calculating quantum mean values in the case of general shallow QAOA circuits. The core of our algorithm is to divide a large QAOA instance into many independent small instances by using methods of graph decomposition and  graph-to-circuit mapping. Since small instances can be handled independently, our algorithm is a natural parallel algorithm.
Analysis and numerical tests show that the run time of our algorithm scales linearly with qubits number for most of optimization problems. For the special QAOA instances that its connections in problem Hamiltonian scales with size of instance, e.g. Sherrington-Kirkpatrick (S-K) model, though the run time scales exponentially, our algorithm still behaves much better than the state-of-the-art tensor network algorithm. Our results definitely show that current NISQ processors has no advantages in quantum mean value problem in QAOA.

\section{The Quantum Approximate Optimization Algorithm}
In the seminal paper of Farhi, Goldstone, and Gutmann, the QAOA is proposed as a variational quantum algorithm to produce approximate solutions for combinatorial optimization (CO) problems \cite{op1}. Since then numerous research works on QAOA have been shown both theoretically \cite{theo1, theo2, theo3, theo4, theo5, theo6, theo7, theo8, theo9, theo10}and experimentally \cite{ex1, ex2, ex3, ex4, ex5, op3}. 
Similar to quantum annealing (QA), in which CO problems are modeled as the form of Ising Hamiltonian, the QAOA also starts with Ising form as cost function. The cost function is quadratic and its general operator form reads    
\begin{equation}
\hat{C}=\sum_{i,j}w_{ij}\hat{Z}_{i}\hat{Z}_{j}+\sum_{k}h_{k}\hat{Z}_{k},
\end{equation}
where $\hat{Z}_{i}$ are Pauli-Z operators with eigenvalues $\pm 1$, $w_{ij}$ and $h_{k}$ represent weights and bias respectively. If we denote computational basis vector as $|z\rangle$ with $z\equiv z_{1}z_{2}...z_{n}$ are $\lbrace 1, -1\rbrace$ bitstrings, the $\hat{C}$ is diagonal in the computational basis vector $|z\rangle$ such that $\langle z|\hat{C}|z\rangle=C(z)$.

The bitstring $z$ that minimize (or maximize) $\langle z|\hat{C}|z\rangle$ is the optimal solution of optimization problems. Since most optimization problems are NP hard, there is still a lack of efficient algorithms. The QAOA, however, approximates the optimal solution by generating quantum state in quantum circuit that as close as possible to the optimal basis vector. This is done by introducing another operator $\hat{B}=\sum_{i}\hat{X}_{i}$ and generating quantum state with $2p$ parameters as
\begin{equation}
|\vec{\gamma}, \vec{\beta}\rangle=\hat{U}_{B}(\beta_{p})\hat{U}_{C}(\gamma_{p})\cdots\hat{U}_{B}(\beta_{1})\hat{U}_{C}(\gamma_{1})|+\rangle,
\end{equation}
where $\hat{U}_{C}(\gamma)=e^{-i\gamma\hat{C}}$, $\hat{U}_{B}(\beta)=e^{-i\beta\hat{B}}$, and the initial state $|+\rangle=\sum_{z}|z\rangle/\sqrt{2^{n}}$ be the uniform superposition of computational basis vectors. The expectation value of cost operator $\hat{C}$ 
\begin{equation}
E_{p}(\vec{\gamma}, \vec{\beta})=\langle\vec{\gamma}, \vec{\beta}|\hat{C}|\vec{\gamma}, \vec{\beta}\rangle
\end{equation}
is then optimized via outer-loop classical optimizer to find optimal parameters $\lbrace\vec{\gamma}_{opt}, \vec{\beta}_{opt}\rbrace$. Once the optimized state $|\vec{\gamma}_{opt}, \vec{\beta}_{opt}\rangle$ is obtained, 
we can perform sampling from the optimized quantum circuit and output the bitstring that minimized the $C(z)$ as approximate solution. 

The core of QAOA is the computation of quantum mean value $E_{p}(\vec{\gamma}, \vec{\beta})$, and there are two different implementations. The first one, which estimates the $E_{p}(\vec{\gamma}, \vec{\beta})$ from sampling results, is realized by making repeated calls to the quantum processor. The alternative one is to calculate the $E_{p}(\vec{\gamma}, \vec{\beta})$ directly on the classical computers. In both cases, constrained by the current capabilities of quantum processors and classical simulations, the computation is limited to the case of shallow depth circuit such that in general $p\le 4$.
In the following we focus on the efficient classical computation of quantum mean values for shallow depth circuit.

\section{Graph decomposition based classical algorithm}
Since the cost operator $\hat{C}$ is consist of independent terms $\hat{Z}_{i}\hat{Z}_{j}$ and $\hat{Z}_{k}$, the $E_{p}(\vec{\gamma}, \vec{\beta})$ can be recast as 
\begin{equation}
E_{p}(\vec{\gamma}, \vec{\beta})=\sum_{ij}w_{ij}\langle\hat{Z}_{i}\hat{Z}_{j}\rangle_{p}+\sum_{k}h_{k}\langle\hat{Z}_{k}\rangle_{p},
\end{equation}
where $\langle\hat{Z}_{i}\hat{Z}_{j}\rangle_{p}\equiv\langle\vec{\gamma},\vec{\beta}|\hat{Z}_{i}\hat{Z}_{j}|\vec{\gamma},\vec{\beta}\rangle$, and the same form for $\langle\hat{Z}_{k}\rangle_{p}$. The computation of $E_{p}(\vec{\gamma},\vec{\beta})$ now depends on the computation of individual terms $\langle\hat{Z}_{i}\hat{Z}_{j}\rangle_{p}$ and $\langle\hat{Z}_{k}\rangle_{p}$. The sum form of Eq. (4) indicates that an efficient parallel algorithm is feasible if we can come up with a procedure to calculate individual terms. This critical procedure is the graph decomposition \cite{networkx} that be introduced below.

It should be noted here that the initial idea of graph decomposition has already been mentioned in the seminal paper of QAOA when dealing with Max-cut problem \cite{op2}. Those authors realized that there exists only finite subgraphs for each edge $(ij)$ in Max-cut graph for finite $p$ and the expectation value $E_{p}(\vec{\gamma}, \vec{\beta})$ is thus determined by expectation values related to subgraphs and the number of occurrences of the subgraphs~\cite{op2}. Surprisingly, no specific algorithm implementation is given in the follow-up study. Besides, there are no further discussions about extending graph decomposition of Max-cut to general combinatorial optimization problems. 
Our contribution here is connecting Ising cost operator $\hat{C}$ of optimization problems to weighted graph representation $G$ and develop specific graph decomposition based algorithm to realize efficient calculations for shallow depth QAOA circuits. 

\begin{algorithm}[H]
    \renewcommand{\algorithmicrequire}{\textbf{Input:}}
    \renewcommand{\algorithmicensure}{\textbf{Output:}}
	\caption{Graph Decomposition Algorithm}
	\begin{algorithmic}[1]
		\Require $\hat{C},\quad p,\quad (\vec{\gamma}, \vec{\beta})$
		\Ensure $E_{p}(\vec{\gamma}, \vec{\beta})$
		\State{weighted graph $\mathcal{G}$ $\gets$ $\hat{C}$}
		\State{{\it subgraph\_set $\gets$ $\varnothing$, \quad  expectation\_set $\gets$ $\varnothing$ }}
		\State{$E_{p}(\vec{\gamma}, \vec{\beta})$ $\gets$ $0$}
		\For{{\it nodes}  and {\it edges} in $\mathcal{G}$}	
		\State{{\it elements} $\gets$ \{ {\it nodes} or {\it edges} \} }	
		\For{ i $\gets$ $1$ to $p$}
		\State{{\it elements} $\gets$ {\it elements} $\cup$ adjacency nodes and edges}
		\EndFor
		\State{ \it{subgraph $\mathcal{G}(i, j, p)$ or $\mathcal{G}(k, p) \gets$ {\it elements}}}
		\State{{\it subgraph\_set} $\gets$ {\it subgraph} $\cup$ {\it subgraph\_set}}
		\EndFor
		
		\For{{\it subgraph} in {\it subgraph\_set}}
			\State{qubits number $N \gets$ node numbers of {\it subgraph}}
			\State{ {\it prepare $|+\rangle ^{\otimes N}$} initial state}
			\For{i $\gets$ 1 to $p$ }
			 	\For{edge in {\it subgraph}}
					\State{$(q_m, q_n)$ $\gets$ nodes of {\it edge}}
					\State{$w_{mn} \gets$ weight of edge}
					\State{apply $RZZ(\gamma_i \times \omega_{mn})$ {\it gate} to $(q_m, q_n)$}
				\EndFor
				\For{{\it node} in {\it subgraph}}
					\State{$q_m, h_k \gets$ $node$, weight of node}
					\State{apply $RZ(\gamma_i \times h_k)$ {\it gate} } to $q_m$
					\State{apply $RX(\beta_i \times h_k$) {\it gate} to $q_m$}
				\EndFor
			\EndFor
		\State{{\it expectation $\gets$ calculating $\langle\hat{Z}_{i}\hat{Z}_{j}\rangle_{p}$ or $\langle\hat{Z}_{k}\rangle_{p}$}}
		\State{{\it expectation\_set} $\gets $ {\it expectation} $\cup$ {\it expectation\_set}}
		\EndFor

		\For{{\it expectation} in {\it expectation\_set}}
		\State{{\it $E_{p}(\vec{\gamma}, \vec{\beta})$ $\gets$ $\sum_{ij}w_{ij}\langle\hat{Z}_{i}\hat{Z}_{j}\rangle_{p}+\sum_{k}h_{k}\langle\hat{Z}_{k}\rangle_{p}$ }}
		\EndFor
	\end{algorithmic}
\end{algorithm}

\begin{figure*}
    \centering
    \includegraphics[scale = 0.4]{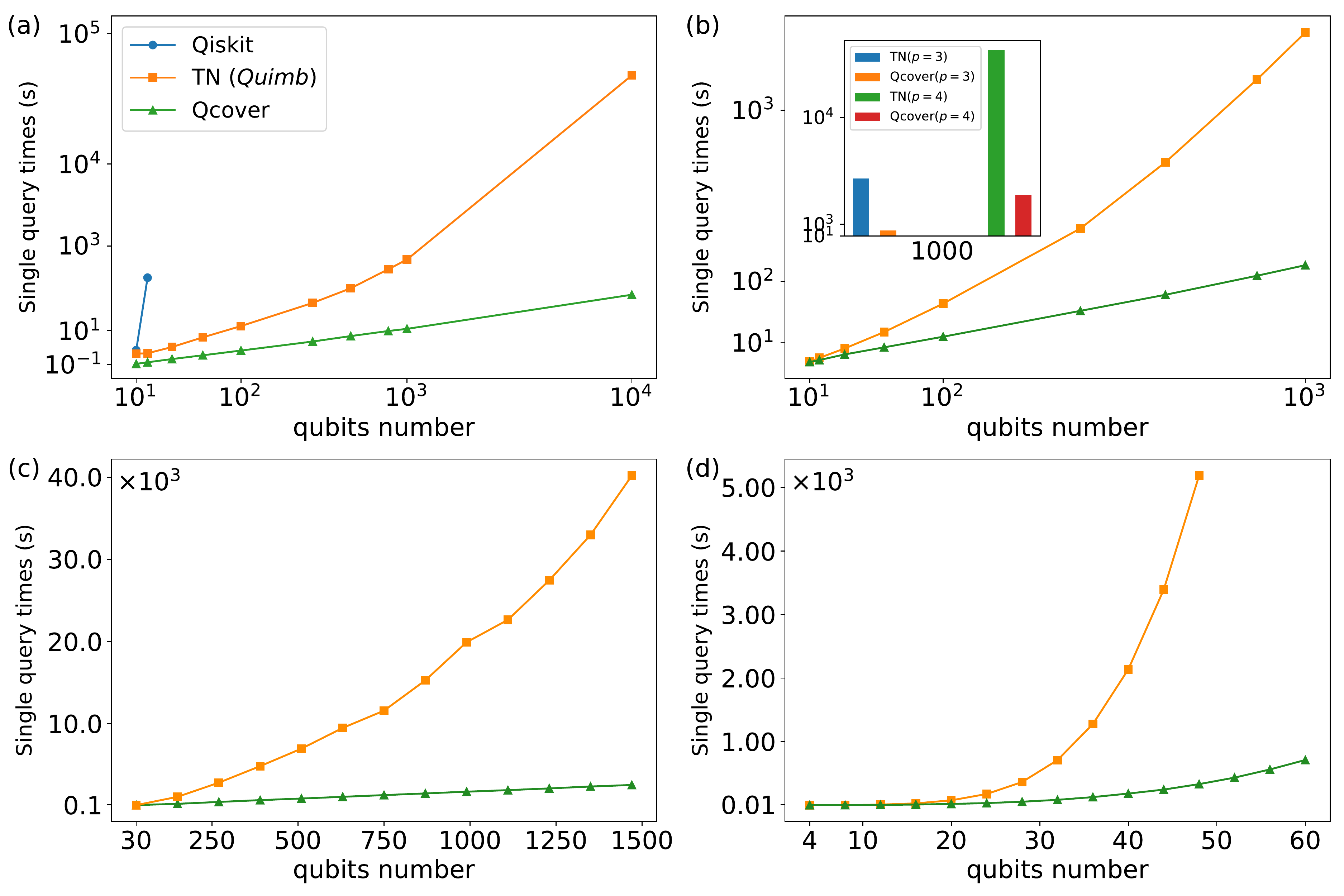}
    \caption{Benchmarking results for random 3-regular Max-cut, graph coloring and S-K model. (a) and (b) show the results of Max-cut, in which $p=1$ in (a) and $p=2$ in (b). For $p=3, 4$, we give the result in the case of fixed $1000$ qubits number. (c) show the result of coloring random generated 3-degree graph with three colors in case of $p=1$. In the case of Max-cut and graph coloring, Qcover scales linearly with qubits number while other methods scale exponentially. Since the graph of S-K model is complete graph, both Qcover and other methods scale exponentially but Qcover behaves much better as can be seen in (d) with $p=1$.    }
    \label{fig:performance}
\end{figure*}

The idea of our graph decomposition is based on two observations. The first one is that any factors in the operators $\hat{U}_{c}(\vec{\gamma}), \hat{U}_{B}(\vec{\beta})$ that do not involve qubits $i$ or $j$ commute through $\hat{Z}_{i}\hat{Z}_{j}, \hat{Z}_{i}$, and can thus be cancelled out without affecting the calculation of $\langle\hat{Z}_{i}\hat{Z}_{j}\rangle_{p}, \langle\hat{Z}_{i}\rangle_{p}$.              The other one is that the cost operator $\hat{C}$ in Eq. (1) has one-to-one correspondence with the weighted graph $G$. The $\hat{Z}_{i}$ corresponds to the node $i$ in $G$, while $w_{ij}$ and $h_{k}$ are the weight of edge $(ij)$ and the bias of node $k$ respectively. From the above two observations, it can be concluded that individual term like $\langle\hat{Z}_{i}\hat{Z}_{j}\rangle_{p}$ depends only on the subgraph $G(i,j,p)$, which involves nodes $i, j$ and other nodes connecting to them with no more than $p$ distance. Also the term like $\langle\hat{Z}_{k}\rangle$ depends on the subgraph $G(k, p)$. 
For each subgraph $G(i, j, p)$ or $G(k, p)$, there exists an one-to-one correspondence with the quantum circuits representation for the computation of $\langle\hat{Z}_{i}\hat{Z}_{j}\rangle_{p}$ or $\langle\hat{Z}_{k}\rangle_{p}$, in which the qubits number equals to nodes number in subgraphs. The edge $(ij)$ with weight $w_{ij}$ in the subgraph corresponds to two-qubit gate $e^{-i\gamma w_{ij}\hat{Z}_{i}\hat{Z}_{j}}$ operating on qubits $i, j$. The node $k$ with bias $h_{k}$ in the subgraph corresponds to one-qubit operation $e^{-i\gamma h_{k}\hat{Z}_{k}}$. 
We give a simple example in Fig. \ref{f1} to illustrate the basic procedure of graph decomposition algorithm.
The size of subgraphs is determined by $p$ and degree of connection of $G$, which is usually independent of $n$ unless the $G$ is a complete graph. The weighted graphs of optimization problems are sparse or medium density in most cases, making the size of subgraphs not so big for small $p$.
For example, in the case of 3-regular Max-cut, the subgraphs for $p=1$ contains only $6$ qubits which can be easily simulated classically, regardless of the size of the problem. Since the computational complexity of graph decomposition is $o(n)$, our algorithm scales linearly with sparse density optimization problems in the case of shallow circuits. The size of subgraphs grows as $p$ increases, which increase the run time for calculation of each individual term in Eq. (4). Suppose that the weighted graph $G$ has an average degree $d=l/n$ with $l, n$ are total number of edges and nodes respectively. If $d$ keeps constant as $n$ grows, e.g., 3-regular Max-cut, then the size of subgraphs is constant for arbitrary finite $p$, which implies the time complexity of our algorithm be $C(d,p)o(n)$ with $C(d, p)$ is a constant depending on the value of $(d, p)$. For large $d$ or $p$, the value of $C(d, p)$ may large enough that beyond current classical simulation but the resource requirement is independent of $n$. For special case that $d$ depends on $n$, e.g., $d=n-1$ in S-K model, $C(d, p)$ now depends on $n$ with exponential complexity in general for current classical algorithms. For most optimization problems, $d$ is constant or $o(\log n)$, the run time of graph decomposition algorithm scales linearly or polynomially with nodes number $n$.

\section{Numerical results}

The pseudo code implementation of graph decomposition algorithm is given in \textbf{Algorithm 1}. The input is the cost operator $\hat{C}$ and $p$, while the output is the expectation value $E_{p}(\vec{\gamma}, \vec{\beta})$. The first step is convert the $\hat{C}$ to corresponding weighted graph $G$, and then decompose the $G$ starting from nodes and edges of $G$ according to the value of $p$. The total number of subgraphs is the sum of numbers of nodes and edges. For each subgraph $G(i, j, p)$ or $G(k, p)$, it can be mapped to a corresponding QAOA quantum circuit for the calculation of expectation value $\langle\hat{Z}_{i}\hat{Z}_{j}\rangle_{p}$ or $\langle\hat{Z}_{k}\rangle_{p}$. The simulation of quantum circuits can be implemented by kinds of method, e.g. statevector or tensor contraction. At last the $E_{p}(\vec{\gamma}, \vec{\beta})$ is obtained by adding every terms $\langle\hat{Z}_{i}\hat{Z}_{j}\rangle_{p}$ or $\langle\hat{Z}_{k}\rangle_{p}$. Based on \textbf{Algorithm 1}, we have developed a quantum software package named {\it QCover} \cite{Qcover} to help finding optimal parameters of shallow QAOA circuits more faster than existing software .

To test our algorithm, we perform numerical experiments on three classical optimization problems, i.e. 3-regular Max-cut, graph coloring and S-K model by comparing the {\it QCover} with the {\it IBM Qiskit} \cite{qiskit} and the {\it Quimb} \cite{quimb}. The {\it IBM Qiskit} uses statevector method to realize QAOA simulation, while the {\it Quimb} takes the method of tensor network contraction. The most important part of tensor method is to search the optimal contraction path. The {\it Quimb} provides many heuristic algorithms to determine contraction path and we choose the best one in our instance tests.  
The engine of the {\it QCover} to compute subgraphs could be statevector or tensor, depending on the characteristics of subgraphs. For example, statevector engine is more efficient than tensor one in the case of 3-regular Max-cut with $p=1$, while it is better to use tensor engine for $p\geq 2$. The {\it QCover} can automatically choose the best engine to complete the computation. The cost operator of Max-cut and S-K are given with the same form
\begin{equation}
\hat{C}=\sum_{i\neq j}w_{ij}\hat{Z}_{i}\hat{Z}_{j},
\end{equation}
where $w_{ij}=1$ for Max-cut and $w_{ij}$ is randomly chosen to be $\pm 1$ for S-K model. For graph coloring problem, we first model it with quadratic unconstrained binary optimization (QUBO) form and then transformed it into Isng cost form $\hat{C}$ \cite{QUBO}. The {\it QCover} has an instance library to automatically generate the weighted graph $G$ corresponding to $\hat{C}$ for above three problems.

The weighted graph $G$ of S-K model is a complete graph \cite{SK}, while it is constant density graph for Max-cut or graph coloring problem \cite{QUBO}. The size of subgraphs for complete graph equals to $G$ even for $p=1$ case, which constrains the calculations for small size. For constant density graph, the size of subgraphs depend only on the density and the value of $p$, which makes us capable of dealing with large size instances given that the size of subgraphs is within the scope of computing power.
The Fig. 2 shows the test results in which horizontal axis represents the node numbers (or qubit numbers) of weighted graph $G$ and longitudinal axis represents the time to complete calculation of $E_{p}(\vec{\gamma}, \vec{\beta})$ with $(\vec{\gamma}, \vec{\beta})$ is given randomly. For each data, we show the average time value of 5 independent tests. The {\it QCover} scales linearly with the node numbers in Max-cut and graph coloring problems as shown in sub-figures (a), (b) and (c) of Fig. 2, while the {\it qiskit} and {\it Quimb} behave exponentially. For S-K model result shown in the sub-figure (d) of Fig. 2, the {\it QCover} also has better performance but scales exponentially due to the size of subgraphs equal to the node numbers. It should be noted here that all date from the {\it QCover} are obtained from the personal computer run without using parallel computing.  Since the subgraphs obtained from weighted graph can be handled individually with diffient computing cores, the {\it QCover} supports high performance parallel computing and better speed-up can be achieved.

\section{Discussion and conclusion}
Although we have shown quantum mean values for shallow QAOA circuits can be efficiently solved by classical computers, it remains an open question for other VQAs, e.g., VQE and QML. Besides, the sampling in the last stage of QAOA, which has been shown classically intractable, still requires NISQ processor. Our algorithm and software, however, can be used as a powerful classical-assisted tool to help finding and realizing the possible optimization problems that is suitable to demonstrate application-level quantum advantage in NISQ processor. It is also helpful of our algorithm for the verification and benchmarking of NISQ computers. Since QAOA deals with Ising Hamiltonian, our algorithm has the potential to assist realizing fast approximate ground state preparation of arbitrary long-range Ising-type Hamiltonian \cite{ex3}.

In summary, we have present a graph decomposition algorithm to realize efficient classical computation of quantum mean values for shallow QAOA circuits. For most optimization problems, the run time of our algorithm scales linearly with the instance size in the case of shallow circuits. When combined with high-performance parallel computing in the next stage, the performance of our algorithm can be further significantly improved. Our algorithm and the related software {\it Qcover}, when used with NISQ processor, may accelerate the demonstration of application-level quantum advantage.

\hfill

\begin{acknowledgments}
\textbf{Acknowledgments:} The authors would like to thank Bai-Ting Liu, Xingyao Wu, Diqing Chen and Yuxuan Wang for valuable discussions. This work is supported by the Beijing Academy of Quantum Information Sciences. Data availability: All codes and data can be found in GitHub https://github.com/BAQIS-Quantum/Qcover.

\end{acknowledgments}

\end{document}